\documentclass[aps,pra,twocolumn,superscriptaddress,nofootinbib,longbibliography]{revtex4-2}

\usepackage{amsmath,amsfonts,amsthm,amssymb}
\usepackage{graphicx}
\usepackage{color}
\usepackage{xurl}
\usepackage{nicefrac}
\usepackage{hyperref}
\usepackage{orcidlink}

\usepackage[english]{babel}

\usepackage{multirow}
\usepackage{textcomp}
\usepackage{array}
\usepackage{booktabs}

\begin{document}

    \title{Deep-Learning-Based Radio-Frequency Side-Channel Attack on Quantum Key Distribution}

    \author{Adomas Baliuka~\orcidlink{0000-0002-7064-8502}}
    \affiliation{Fakult\"at f\"ur Physik, Ludwig-Maximilians-Universit\"at, 80799 M\"unchen, Germany}
    \affiliation{Munich Center for Quantum Science and Technology, 80799 M\"unchen, Germany}

    \author{Markus Stöcker~\orcidlink{0009-0005-3515-8419}}
    \affiliation{Fakult\"at f\"ur Physik, Ludwig-Maximilians-Universit\"at, 80799 M\"unchen, Germany}
    \affiliation{Munich Center for Quantum Science and Technology, 80799 M\"unchen, Germany}

    \author{Michael Auer~\orcidlink{0009-0000-2480-0991}}
    \affiliation{Fakult\"at f\"ur Physik, Ludwig-Maximilians-Universit\"at, 80799 M\"unchen, Germany}
    \affiliation{Munich Center for Quantum Science and Technology, 80799 M\"unchen, Germany}
    \affiliation{Universität der Bundeswehr, 85577 Neubiberg, Germany}

    \author{Peter Freiwang}
    \affiliation{Fakult\"at f\"ur Physik, Ludwig-Maximilians-Universit\"at, 80799 M\"unchen, Germany}
    \affiliation{Munich Center for Quantum Science and Technology, 80799 M\"unchen, Germany}

    \author{Harald Weinfurter~\orcidlink{0000-0001-6882-3909}}
\email{h.w@lmu.de}
    \affiliation{Fakult\"at f\"ur Physik, Ludwig-Maximilians-Universit\"at, 80799 M\"unchen, Germany}
    \affiliation{Munich Center for Quantum Science and Technology, 80799 M\"unchen, Germany}
    \affiliation{Max-Planck-Institut f\"ur Quantenoptik, 85748 Garching, Germany}
    \affiliation{Institute of Theoretical Physics and Astrophysics, Faculty of Mathematics, Physics, and Informatics, University of Gdańsk, 80-308 Gdańsk, Poland}

    \author{Lukas Knips~\orcidlink{0000-0002-7404-1708}}
\email{lukas.knips@mpq.mpg.de}
    \affiliation{Fakult\"at f\"ur Physik, Ludwig-Maximilians-Universit\"at, 80799 M\"unchen, Germany}
    \affiliation{Munich Center for Quantum Science and Technology, 80799 M\"unchen, Germany}
    \affiliation{Max-Planck-Institut f\"ur Quantenoptik, 85748 Garching, Germany}

    \begin{abstract}
        Quantum key distribution (QKD) protocols are proven secure based on fundamental physical laws, however, the proofs consider a well-defined setting and encoding of the sent quantum signals only.
        Side channels, where the encoded quantum state is correlated with properties of other degrees of freedom of the quantum channel, allow an eavesdropper to obtain information unnoticeably as demonstrated in a number of hacking attacks on the quantum channel.
        Yet, also classical radiation emitted by the devices may be correlated, leaking information on the potential key, especially when combined with novel data analysis methods.

        We here demonstrate a side-channel attack using a deep convolutional neural network to analyze the recorded classical, radio-frequency electromagnetic emissions.
        Even at a distance of a few centimeters from the electronics of a QKD sender employing frequently used electronic components we are able to recover virtually all information about the secret key.
        Yet, as shown here, countermeasures can enable a significant reduction of both the emissions and the amount of secret key information leaked to the attacker.
        Our analysis methods are independent of the actual device and thus provide a starting point for assessing the presence of classical side channels in QKD devices.
    \end{abstract}

    \maketitle

    \section{Introduction}

    Quantum Key Distribution (QKD)~\cite{bennett_quantum_1984,gisin_quantum_2002,Scarani2009,Diamanti2016,pirandola_advances_2020,xu_secure_2020} is one of the most mature quantum technologies.
	It allows two authenticated parties to use a quantum channel to exchange a cryptographic secret, which they can later use for symmetric cryptography.
    Using fundamental physical principles, QKD allows to quantify the amount of information leakage to an eavesdropper and to subsequently eliminate it entirely using appropriate postprocessing.
        QKD is used both for short-distance and long-distance communication via free-space~\cite{SchmittManderbach2007,Nauerth2013,Handheld2022,Micius2022,Knips22} and fiber-based~\cite{Jouguet2013,Yin2016,Boaron2018,Zhang2018} links with first or planned implementations in large networks~\cite{Chen2021, EUQCITechnicalBackground}.
	With a plethora of testbeds and implementations in multinational industry-oriented consortia, QKD has reached commercial end-user availability.

	Yet, despite its conceptual elegance, the practical security hinges upon the quality of the implementation, in particular, strict adherence to the theoretical model used to prove security.
	The quantum states sent over the quantum channel have to be prepared precisely within the requirements of the QKD protocol.
    Any correlation with any other degree of freedom, but also with classical properties of the devices used, potentially opens side channels~\cite{xu_secure_2020,pirandola_advances_2020, PhysRevA.78.042333, Lydersen2010}.
    These will allow an eavesdropper to infer the key by measurements unnoticeable to the users.
    We refer to side channels exploited by interacting with the quantum channel as \textit{quantum side channels} and all other side channels as \textit{classical side channels}.

    Electronic devices continually emit electromagnetic radiation and are in turn influenced by it.
    Thus, the operation of security critical devices may be influenced by \textit{active attacks}~\cite{Giechaskiel2020} rendering them insecure, such as demonstrated on quantum random number generators~\cite{Smith2021}.
    However, if emissions from a device are correlated with sensitive information processed by it, a critical side channel opens up widely and allows much simpler \textit{passive attacks}.
    They do not need any manipulation of device components and are practically impossible to detect.
    Investigations of information leakage from conventional communication systems via electromagnetic radiation go back to at least the 1940s, later under the US military codename TEMPEST~\cite{nsa_tempest_1972}.
    TEMPEST attacks now refer to eavesdropping via electromagnetic or acoustic side channels and are widely considered in security specifications and during certification of security critical systems.
	Technologies such as software-defined radio (SDR), specialized probes~\cite{heyszl_localized_2012} and particularly deep learning~\cite{masure_comprehensive_2019,duan_machine_2021} make the exploitation of vulnerabilities much easier and more effective.

    In reaction to quantum hacking attacks on QKD devices countermeasures have been developed to protect and secure against side channel attacks on the quantum channel~\cite{xu_secure_2020}.
%	However, as QKD devices rely on many electronic components used in standard security-critical systems, they are expected to be vulnerable also to classical side channel attacks, e.g., based on electromagnetic emissions.
  Yet, as standard electronic components, especially logic units such as FPGAs, ASICs or CPUs emit electromagnetic radiation, they also open a new, classical side channel for attacks on potentially every QKD system.

 	Here, we demonstrate a deep-learning-based side-channel attack on a QKD device using radio-frequency (rf) emissions at frequencies up to a few GHz.
    Our setup does not require expensive specialized equipment and works with few computational resources.
    In some scenarios, our attack is able to recover virtually all information about the secret key.
    In contrast to a recent attack on QKD single photon detector electronics~\cite{Durak2022}, our attack targets the control electronics in general.
    We demonstrate its power and security threat on a QKD sender module.
    We analyze how the information leakage depends on distance to the device, as well as to which extent it can be mitigated with improved design and shielding.
    The QKD sender electronics inspected here is home-built, but, since it is made from conventional electronic components also used in other QKD systems, it is representative for other, also commercial systems.
    In addition, our data evaluation may also be applied to attack via other weak points, e.g., power consumption~\cite{PARK202136} or acoustic side channels~\cite{genkin_acoustic_2017}.
    This clearly demonstrates that a detailed examination of classical side channels is important for future QKD devices and networks.
    Emission security should be considered from the early design stages~\cite{ott2009electromagnetic} until the deployment of QKD devices.

    \section{Experimental setup and data collection}\label{sec:experimental-setup}

    \subsection{Sender module and attacker setup}\label{subsec:attacker-setup}
    \begin{figure}[htb]
        \centering
        \includegraphics[width=0.45\textwidth]{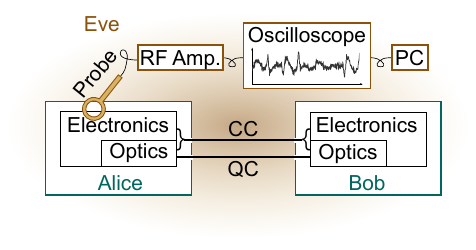}
        \caption{Sender (Alice) and receiver (Bob) devices, comprising of electronics and optics, are connected via a quantum channel (QC) and a classical channel (CC).
		The eavesdropper (Eve) has access to both channels.
		Eve measures Alice's emissions using a near-field probe for magnetic fields or a log-periodic antenna (not shown) for far-field measurements, whose radio-frequency (RF) signal is amplified, captured by an oscilloscope and evaluated on a PC~\cite{devices}.
		}
        \label{fig:attackersetup}
    \end{figure}

	The sender module is a home-built BB84 polarization-encoding decoy-capable QKD device building upon the device presented in~\cite{Handheld2022}.
    It features a field-programmable gate array (FPGA), which controls four distinct vertical-cavity surface-emitting laser (VCSEL) drivers~\cite{Auer2020,Auer21,devices}.
	The drivers are connected to four VCSELs emitting short light pulses with a wavelength around $850\,\mathrm{nm}$, which are subsequently polarized by differently rotated polarization filters.
	This way, the sender device can emit optical pulses with either of the four polarization directions (horizontal/H, vertical/V, diagonal/P, anti-diagonal/M).
	For the measurements presented here, the module is sending random streams of \textit{symbols} (H, V, P, M) at a symbol rate of $f_{\mathrm{clk}}=100\,\mathrm{MHz}$.

	For our attack, we record electromagnetic near-field emissions from the printed circuit board (PCB) of the QKD sender using a magnetic near-field probe, an rf amplifier and an oscilloscope~\cite{devices} with a bandwidth of $8\,\mathrm{GHz}$, see Fig.~\ref{fig:attackersetup}.
    Given the symbol rate of $100\,\mathrm{MHz}$, we sample the emissions signal at $f_{\mathrm{samp}}=10\,\mathrm{GSa/s}$ to obtain $100$ voltage samples per symbol.
    The oscilloscope memory sets the length of the total time trace to $N_{\mathrm{meas}} = 2\,\mathrm{MSa}$, corresponding to a measurement duration of $t_{\mathrm{seq}}=200\,\mathrm{\mu s}$.
    We hence have sequences of length $N_{\mathrm{seq}}=t_{\mathrm{seq}}\times f_{\mathrm{clk}}=20\,000$ symbols sent by the sender.
	Besides near-field emissions, we also record far-field emissions using a commercially available directed wideband log-periodic dipole antenna.

    \subsection{Near-field spectrum}
    \begin{figure}[htb]
        \centering
        \includegraphics[width=0.45\textwidth]{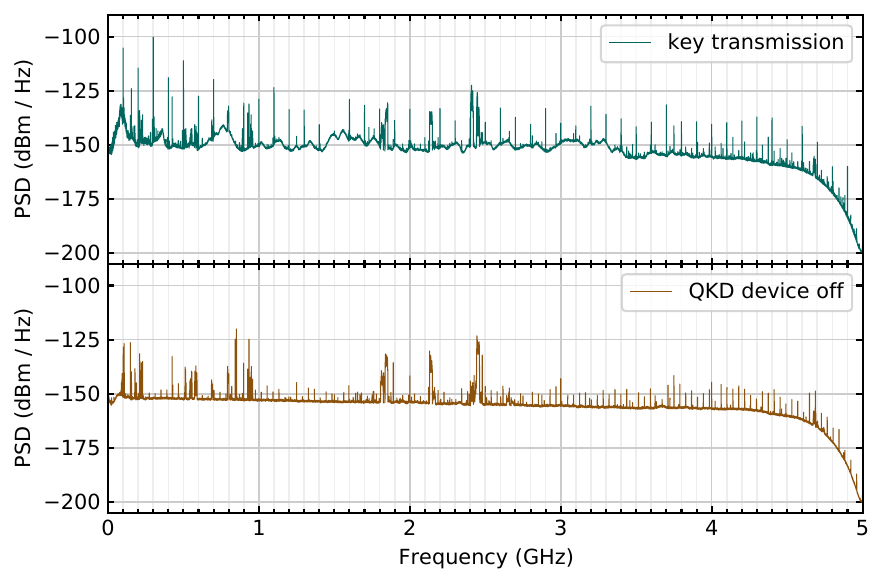}
        \caption{Near-field spectra (power spectral density) of the emissions during a key transmission (green) and during a null measurement (brown) where the QKD sender is not powered.
		The regular spikes on the fine grid in the upper plot are harmonics of the $100\,\mathrm{MHz}$ clock frequency. Frequencies higher than about $5\,\mathrm{GHz}$ are suppressed due to the limited bandwidths of probe, amplifier, and oscilloscope.
        The spectra are obtained by Barlett's method (segment length $100\,000$ samples) and averaged over 30 independent measurements.
        At some frequencies the background exceeds the signal due to the noisy office environment changing in time.
		}
        \label{fig:nearfieldspectrum}
    \end{figure}

	The spectrum of the recorded emissions contains a significant signal at the clock frequency $f_{\mathrm{clk}}$, see Fig.~\ref{fig:nearfieldspectrum}.
	Due to the measurement taking place in a noisy office environment, emissions in communication bands (e.g.\ Wi-Fi, UMTS) contribute to the measured signal.
    Since the deep learning methods used in our attack deal well with such noisy signals, we make no attempt to remove the background noise and use no manual filtering in addition to what naturally occurs in the measurement devices (probe, amplifier, and oscilloscope).
    Our methodology clearly has the advantage that it can cope with standard environments typical, eg.,  for server rooms.
    This makes the attack scenario more realistic and enables device evaluation without the need for specialized shielded facilities.

    \subsection{Single and averaged time traces}

    \begin{figure*}[htb!]
        \centering
        \includegraphics[width=0.95\textwidth]{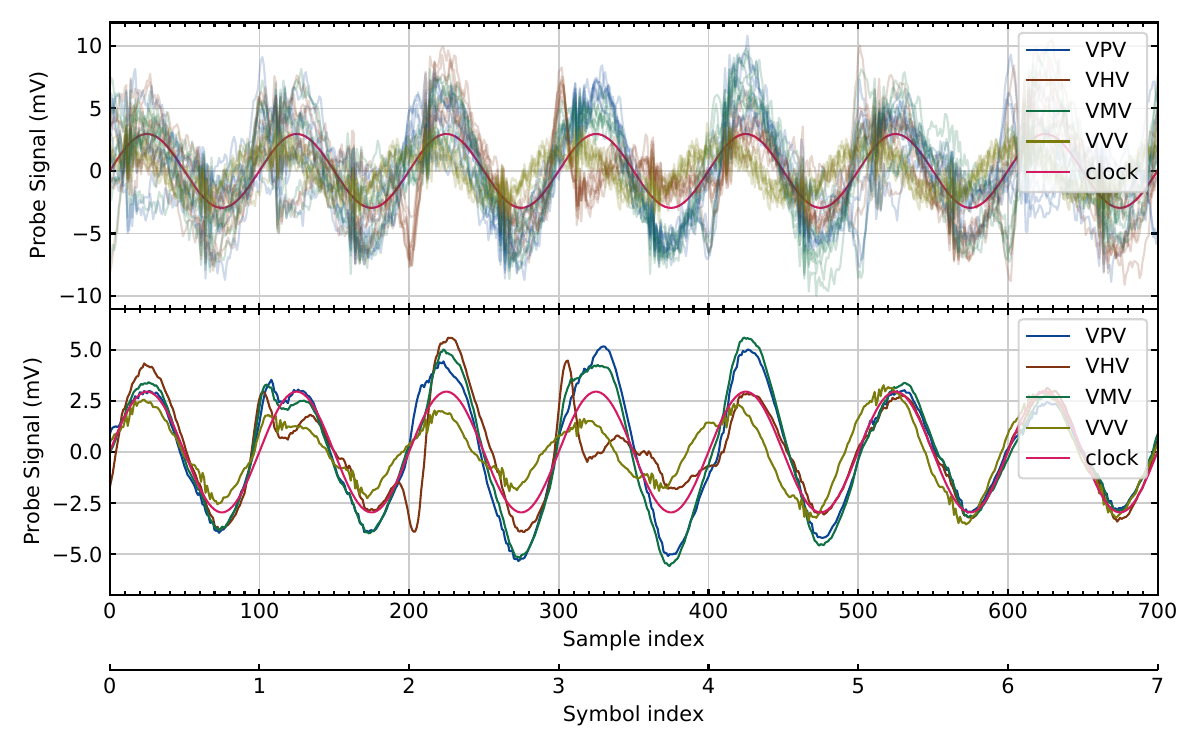}
        \caption{
        Top: For each of the three-symbol key excerpts (VHV, VVV, VPV or VMV), non-overlapping snippets of the recorded time trace (for clarity, only seven per excerpt).
            The range between sample index $200$ and $500$ corresponds to the three symbols in the key excerpt.
            The regions before and after correspond to random symbols which happened to be adjacent to the selected occurrences of the excerpts in the key.
            For reference, the $100\,\mathrm{MHz}$ clock signal is shown, as obtained digitally from the probe signal using a band pass filter.
        Bottom: Averages of all matching snippets (about 300 each) for each symbol combination across one measurement.
            In the regions where random symbols contribute to the average (roughly sample ranges $0\mbox{--}200$ and $500\mbox{--}700$), the differences cancel and the result is close to the clock signal.
		}
		\label{fig:timetraces_combined}
    \end{figure*}

    Let us first examine the data in the time domain.
    For that, we measure a time trace using the near-field probe while the device is repeating a fixed pseudo-random sequence of $20\,000$ symbols.
    Time synchronization between symbols in the key and the measured time trace is achieved using the phase of the clock signal, which is digitally extracted from the emissions, as well as a separately recorded trigger signal signifying the time of the first symbol in the key.
    We verified that the measurement of the trigger signal does not influence the performance of our attack, see App.~\ref{sec:appSynchronization}.

    The near-field probe signal is split into snippets with a length corresponding to a few symbols.
    Since the electronic processes that produce the symbol, especially in the FPGA, take several clock cycles, we thereby make sure to capture all relevant information.
    This yields a set of snippets of the time trace together with the respective sub-sequence of the key.
    Here, for illustration, we choose a snippet length of seven symbols.

    To get more insight, consider, e.g.,  the sub-sequences matching the pattern ``??VXV??'', where ``?'' can be any symbol.
    We group them according to the center symbol ``X'' and show their respective snippets in Fig.~\ref{fig:timetraces_combined}.
    Precise features of these individual time traces are difficult to identify and finding the symbol sequence corresponding to a given time trace by naked eye seems hard.
	However, a first view of the raw data reveals common features and suggests that changes and specific patterns in the measured magnetic field amplitudes correspond to switching between different adjacent symbols, rather than just the symbols themselves.
	When the signals corresponding to the same sub-sequence are averaged (Fig.~\ref{fig:timetraces_combined} bottom), the four different averages differ significantly more around the varying center symbol than at the outermost symbols, where they roughly reproduce the clock signal.

    \section{Machine learning-based attack}	\label{sec:mlattacks}

    Conventional methods to extract confidential information from such emission data require both specialized knowledge of signal processing and detailed models of the emissions, limiting the relevance of the resulting attacks to certain domains of application and specific types of devices or electronic components.
    In contrast, when using machine learning (ML) techniques~\cite{Patterson2017,Schmidhuber2015,Alzubaidi2021}, there is no need to understand how the emissions arise.
    Rather, an effective statistical model of the phenomena is created from recorded data by a training procedure, making the approach more general and adaptable.
	Since we are able to collect training data in a known, controlled environment, we apply \textit{supervised learning}, as opposed to, e.g., \textit{unsupervised learning}~\cite{Celebi2016} or \textit{reinforcement learning}~\cite{Sutton2018}, which may be promising for other types of attacks.

    The attacker's task is mapping a one-dimensional time-series (the recorded emissions) to a sequence of symbols (the raw key).
    To solve it, general sequence-to-sequence methods could be used, such as transformer neural networks~\cite{Xu2022}.
    However, the task can be simplified further by assuming that there are no significant long-time correlations between electromagnetic emissions and the symbol sequence, i.e., that the emissions at a given time only depend on the symbols currently being processed but not on all symbols processed at earlier times.
    Note that the presence of such correlations in the behaviour of the electronics could indicate serious security problems of the QKD device~\cite{Pereira_QKD_corr_sources, yoshino_quantum_2018}.

    With this assumption, it suffices to be able map a short snippet of the time trace to the symbol sent during that time.
    Applying the mapping individually to each snippet then yields the entire key.
    The attack thus becomes a classification task, i.e., mapping a one-dimensional fixed-length time-series to one of four classes (H, V, P, M).
    To solve it, we design and train a convolutional neural network.
    A snippet length of $500$ samples (corresponding to five symbols) as input to the neural network proved sufficient for our attack.
    When predicting the center symbol, this length ensures that all relevant information is contained in the snippet while lowering demands on the precision needed in synchronizing the time trace with the symbol sequence.
    The attack also works with shorter snippet lengths but performs slightly worse.

    \subsection{Attacker model}\label{subsec:attacker-model}
    \begin{figure}[htb!]
        \centering
        \includegraphics[width=0.45\textwidth]{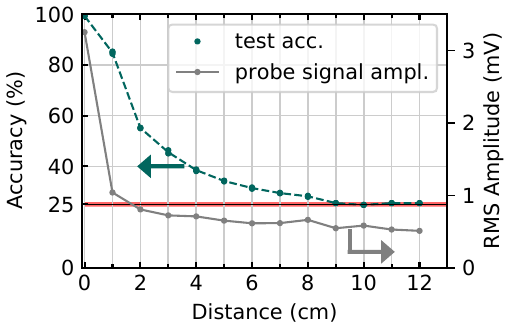}
        \caption{
		Varying distance from the circuit board at a location above the FPGA, which promises high accuracy as indicated in Fig.~\ref{fig:xyscan}a.
		The test accuracy is shown as the average (green dotted line) of three independent attacks (green dots) at each distance.
        For short distances, the test accuracy is remarkably high (about 99\%).
		The baseline is $25\%$, corresponding to randomly guessing one of the four symbols.
        The red area indicates three standard deviations around random guessing, assuming $20\,000$ trials of a $25\%$ Bernoulli distribution.
        The RMS amplitude of the recorded emissions is shown for reference.
        }
        \label{fig:distance}
    \end{figure}
    Our experimental design and data evaluation are motivated by so-called \textit{profiled attacks}~\cite{kim_make_2019}.
    For such attacks, the attacker prepares for the actual attack while having full access to a copy of the victim's device.
    This assumption is in accordance with Kerckhoffs's principle~\cite{Kerckhoffs1883} that security of a system should not depend on secrecy of its design, and is thus appropriate for QKD devices.

    A profiled attack consists of two phases.
    First, in the so-called \textit{profiling}/\textit{training} phase, the attacker uses the copy of the target device and records data corresponding to known symbol sequences chosen at will.
    The data is used to create a model which captures the correlations between secret information and side channels.
    In our case, this so-called \textit{training dataset} consists of a sequence of key symbols, say, $\left({y}^\mathrm{train}_1, {y}^\mathrm{train}_2, \dots, {y}^\mathrm{train}_{N_\mathrm{train}}\right)$ and a sequence of time trace snippets, say, $\left({x}^\mathrm{train}_1, {x}^\mathrm{train}_2,\dots, {x}^\mathrm{train}_{N_\mathrm{train}} \right)$, where ${y}^\mathrm{train}_i$ is the symbol sent during the middle of the snippet ${x}^\mathrm{train}_i$.

    The neural network $f_\theta$, given trainable parameters $\theta$, maps any snippet ${x}$ to the network's \textit{prediction} $f_{\theta}(x) \in \{H,V,P,M\}$.
    The training dataset is used to obtain optimized parameters $\tilde\theta$ such that the model's predictions
    $\left(f_{\tilde\theta}({x}^\mathrm{train}_{1}), f_{\tilde\theta}({x}^\mathrm{train}_{2}), \dots, f_{\tilde\theta}({x}^\mathrm{train}_{N_{\mathrm{train}}})\right)$ approximate the true key symbols $\left({y}^\mathrm{train}_{1}, {y}^\mathrm{train}_{2}, \dots, {y}^\mathrm{train}_{N_{\mathrm{train}}}\right)$.

    In the second, so-called \textit{attack}/\textit{test} phase of the profiled attack, the attacker performs a measurement upon the victim's device during its normal operation, i.e., where the attacker has no control or access to any information except the recorded emissions.
    The attacker records a \textit{test dataset} comprising of recorded emissions only, say, $\left({x}^\mathrm{test}_1, {x}^\mathrm{test}_2, \dots, {x}^\mathrm{test}_{N_\mathrm{test}}\right)$ and obtains an estimate of the key as the predictions of the previously trained model.

    To evaluate the success of the attack, we make use of our access to the true sequence of sent symbols $\left({y}^\mathrm{test}_1, {y}^\mathrm{test}_2, \dots, {y}^\mathrm{test}_{N_\mathrm{test}}\right)$ and compare it to the predictions of the fully trained network $\left(f_{\tilde\theta}({x}^\mathrm{test}_{1}), f_{\tilde\theta}({x}^\mathrm{test}_{2}), \dots, f_{\tilde\theta}({x}^\mathrm{test}_{N_\mathrm{test}})\right)$ by defining the \textit{prediction accuracy} of the test dataset, or simply the \textit{test accuracy}
    \begin{equation}\label{eq:accuracy_equation}
        A = \frac{N_{\mathrm{correct}}}{N_\mathrm{test}},
    \end{equation}
    where $N_{\mathrm{correct}} = \left| \{ i : f_{\tilde\theta}({x}^\mathrm{test}_i) = {y}^\mathrm{test}_{i} \} \right|$ is the number of symbols correctly predicted by the neural network.
    Since the four unique symbols are equally likely to occur in the key, random guessing gives a prediction accuracy of $25\%$.
    We consider an attack successful (in extracting above zero information about the key) if the test accuracy exceeds random guessing by more than three standard deviations of the binomial distribution.
    For a success probability of $25\%$ and $20\,000$ trials, this implies accuracies should be above $25.92\%$.

    We use prediction accuracy because it is intuitive and allows to evaluate the attack on the sender module alone without having to discuss sifting or basis choice.
    A high prediction accuracy implies a successful attack.
    However, prediction accuracy does not directly correspond to the amount of secret information gained by the attacker.
    Even a small but above random prediction accuracy may still allow a critical attack (see App.~\ref{sec:selective_channel_blocking}).
    How the raw key symbol prediction accuracy relates to information leakage is further discussed in App.~\ref{sec:appBitprediction}.

    Note that for a QKD device normal operation implies that the secret key is used only once.
    This means that the attacker has access to only a single time trace of the emissions to extract information about the key.
    While this makes the attacker's task more difficult, we adhere to this restriction, resulting in a \textit{single trace attack}.

    \subsection{Neural network architecture and training}

    Our neural network architecture consists of fully connected layers, one-dimensional convolutions, max pooling, and batch normalization to process traces of the emissions in the time domain.
    To make better use of the data, we employ data augmentation.
    For more details on the architecture of the network and the data augmentation, see App.~\ref{sec:appNetworkTraining}.

    Training state-of-the-art neural networks can require very large datasets and is typically performed on graphics processing units (GPUs) or tensor processing units (TPUs) due to the large computational resources required.
    Since our model is rather small and operates on one-dimensional data, training on a standard laptop with GPU support only takes a few minutes.
	As moving the probe to a new location and performing the measurement takes only a few seconds, our method allows to identify vulnerable components almost in real time.

    \section{Results}

    \subsection{Near-field measurements}\label{subsec:near-field-measurements}

    \begin{figure}[htb!]
        \centering
        \includegraphics[width=0.45\textwidth]{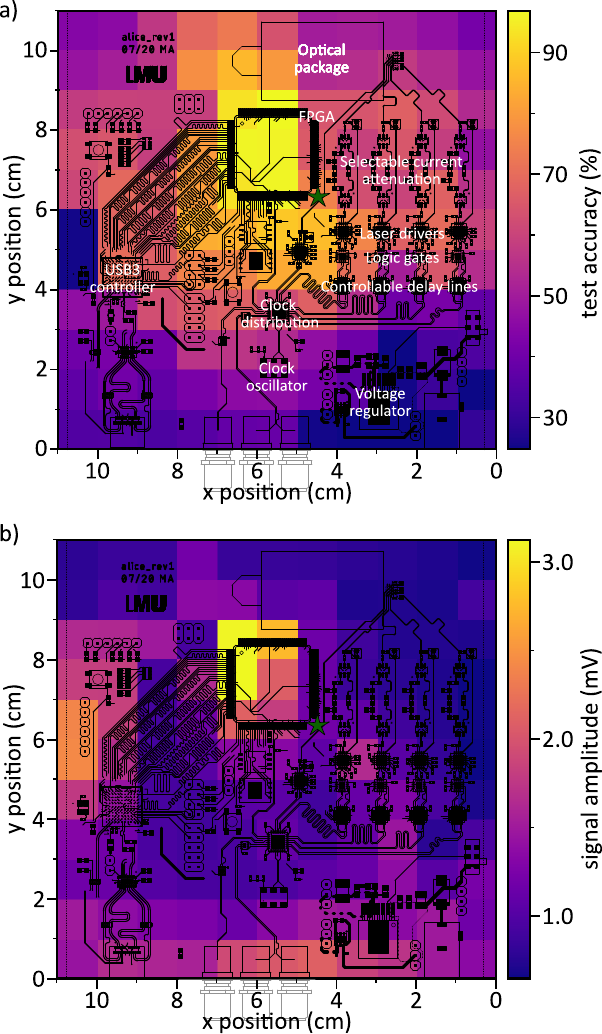}
        \caption{a) Test accuracy of our neural network when trained and tested at respective positions and b) RMS amplitude of recorded emissions.
		Note that both power and accuracy also depend on the angular orientation of the near-field probe, i.e., its rotation about the axis perpendicular to the board, which has been kept fixed.
        A green star indicates the location used for the distance measurement (Fig.~\ref{fig:distance}).
		}
        \label{fig:xyscan}
    \end{figure}

    \begin{figure*}[htb!]
        \centering
        \includegraphics[width=0.95\textwidth]{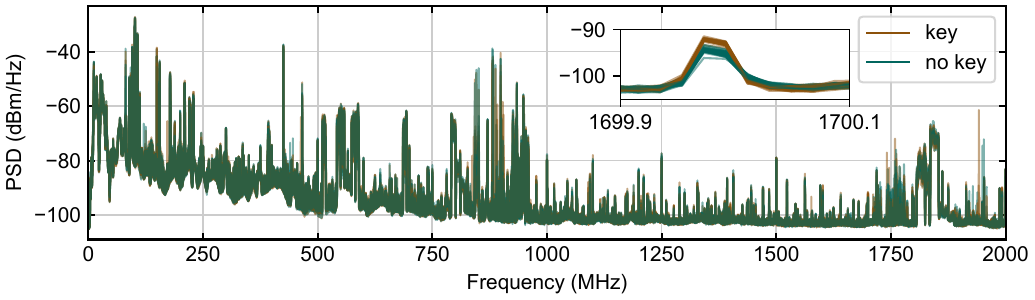}
        \caption{Measurements with an antenna at a distance of about $2.5\,\mathrm{m}$.
		The spectra of $30$ measurement runs when Alice is sending a random key (brown) and not sending a key (green) are clearly distinguishable as shown in the selected signal range around $1.7\,\mathrm{GHz}$ (inset).
        This is even despite various strong noise contributions from Wi-Fi, GPRS, UMTS, Bluetooth, etc.
        }
        \label{fig:antenna}
    \end{figure*}

    We perform the measurement procedure described in Sec.~\ref{subsec:attacker-setup} for various locations of the magnetic near-field probe and collect independent datasets for each location as described in Sec.~\ref{subsec:attacker-model}.
    Two raw keys (each of length $20\,000$) are created by a pseudo-random number generator with different seeds, such that all four symbols are equally likely.
    One of those is used for the training, the other one for the test measurements.
    This is crucial to avoid overfitting~\cite{hastie_elements_2009} and misinterpretation of results.

    Although a single time trace, i.e., the recorded emissions while sending $20\,000$ symbols, is sufficient to demonstrate information leakage, we combine several measurements to increase the amount of training data and thus improve attack performance.
    Trading off longer measurement time for attack performance (see App.~\ref{sec:appNetworkTraining}), each training dataset contains snippets obtained from seven combined time traces (equivalent to a symbol sequence of $140\,000$ symbols), unless indicated otherwise.
    To monitor how much the results depend on unrelated classical communication and background fluctuations in the noisy office environment, we also record each test dataset three times.
    The test datasets are not combined but instead evaluated separately and independently, thus meeting the requirements for a single-trace attack.

    As the first step, we investigate which components and areas of the circuit board contribute to rf emissions or leak information about the key.
    We measure at locations given by a 2-D grid spaced at $10\,\mathrm{mm}$ in $x$ and $y$ directions while keeping the magnetic near field probe at a fixed distance of $10\,\mathrm{mm}$ from the board.
	As shown in Fig.~\ref{fig:xyscan}a, especially measurements close to the FPGA allow to retrieve the key with high accuracy.
	Other components, such as the voltage regulator, also produce significant rf emissions, which, however, are not correlated with the symbols and effectively reduce the attacker's signal-to-noise ratio, leading to lower test accuracy at those locations.
    On the other hand, we obtain high test accuracies also in regions with small amplitudes of recorded emissions (Fig.~\ref{fig:xyscan}b).
    This shows that the test accuracy cannot be inferred from the amplitude of recorded emissions.

    As the second step, we investigate how the distance of the probe from the circuit board affects the accuracy of our attack.
	We select a location (see Fig.~\ref{fig:xyscan}) close to the FPGA, which promises a successful attack.
    Positioning the probe above this location at various distances from the board, we observe a decrease of the test accuracy with increased distance as shown in Fig.~\ref{fig:distance}.
    Yet, the accuracy is above the random guessing value up to distances of $8\,\mathrm{cm}$.

    \begin{figure}[htb!]
        \centering
        \includegraphics[width=0.45\textwidth]{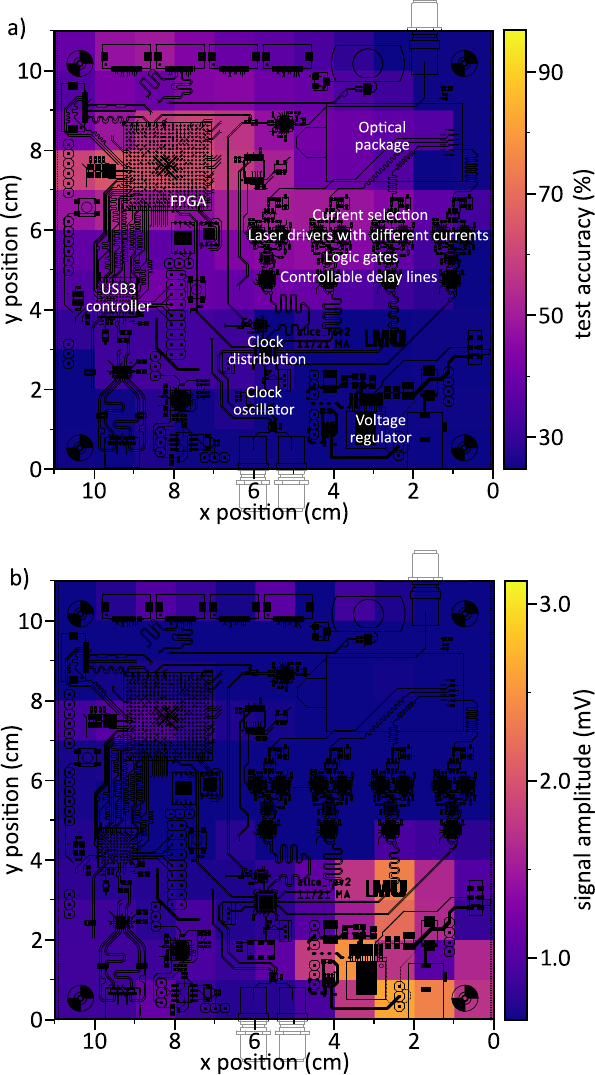}
        \caption{Test accuracy (a) and amplitude of the probe signal (b) of our revised electronics.
        Both accuracy and amplitude are significantly reduced compared to the original electronics as shown in Fig.~\ref{fig:xyscan}.
        The strongest emissions are observed from the voltage regulator (bottom right on the board), which, however, do not carry information about the key.
        Nevertheless, despite the much lower amplitude of emissions from the FPGA, it leaks a significant amount of information about the key.
        }
        \label{fig:xyscan2}
    \end{figure}

    \subsection{Far-field measurements}

    At distances larger than a few centimeters, the near-field probe is no longer effective.
	To investigate whether emissions can still be detected at very long distance, we use a log-periodic dipole antenna~\cite{devices}.
    Due to high background noise in the environment and non-ideal antenna characteristics, we are not able to extract key symbols using the neural network.
    Thus, we pursue the more modest goal of investigating if any emissions at all are present and whether they contain non-zero information about the operation of the QKD device.
    To demonstrate non-zero information, it is sufficient to use emissions to reliably and consistently distinguish two modes of operation of the device.
    To show this, we record about $500$ emission spectra of our unshielded sender device at a distance of about $2.5\,\mathrm{m}$ for two different modes of operation of the QKD sender: sending a random key (``key''), or being turned on but idle (``no key'').
    To exclude influence of background variations in time, the two modes are alternated many times during data collection within datasets.
	By studying the spectra of a training dataset ($396$ spectra), we manually identify a spectral region with a peak which seems highly correlated with the mode of operation (see Fig.~\ref{fig:antenna}).
	Using the selected frequency interval around $1.7\,\mathrm{GHz}$, different machine-learning approaches (support-vector classification, $K$ neighbors classification, linear discriminant analysis) can clearly distinguish those modes of operation (test accuracy of $100\%$ for a test dataset of $94$ spectra).

	Although we are not able to reconstruct the key with this equipment and analysis, this result indicates the possibility of information leakage also over larger distances~\cite{wang_far_2020}.

    \section{Countermeasures}

    There are numerous design and shielding techniques for reducing emissions and preventing information leakage via rf emissions~\cite{Paul2005,EmiShielding2020,ott2009electromagnetic}.
    The following countermeasures significantly reduced emissions from a revised version of our electronics, thus making the attack much less effective (Fig.~\ref{fig:xyscan2}), resulting in the attack being no longer successful for distances larger than $5\,\mathrm{cm}$ (compared to $9\,\mathrm{cm}$ of the former revision as shown in Fig.~\ref{fig:distance}).

    An FPGA in a ball-grid array footprint has been chosen with proper care of differential signal routing, grounding and placing of decoupling capacitors.
	Critical signals have been routed in layers shielded by a ground and a supply-voltage plane.
	An optimization of the FPGA design has not been done, but could further lower the emissions.

    With the addition of metallic shielding with a thickness of a few millimeters, our attack could no longer perform better than random guessing.
    However, note that for a QKD device, an optical channel design with a large puncture of the shielding could significantly reduce the shielding effectiveness~\cite{Robinson1998}.
    We were able to detect a small amount of emissions in front of a hole in the shielding (about $2\times2\,\mathrm{cm}^2$), which allowed to predict key symbols with a test accuracy of more than $27\%$ (highly significant for a key length of $20\,000$ symbols).
    Also, metallic shielding does not help against low frequency magnetic fields~\cite{guri_odini_2018}.

    \section{Conclusion and outlook}
    We have demonstrated that an eavesdropping attack analyzing the electromagnetic emissions from the QKD sender using machine learning can retrieve all information about the key.
    Although we focused our analysis to rf emissions, our methodology and machine learning techniques can also be used for studying information leakage via other potential side channels.
    As shown, countermeasures can reduce the success rate of the attack, yet, they may be difficult to implement, especially if standard electronic components turn out to be the strongest source of information leakage.
    Even small changes in device design or operation environment can have a large effect.
    Since countermeasures are much easier to plan and implement in early design stages of devices, preliminary testing of emissions can be very valuable.

    We want to emphasize the need to test QKD devices and examine information leakage not only via attacks on the quantum channel but also via classical side channels, e.g., electromagnetic emissions, acoustic vibrations, classical message timing and power consumption.
    The methodology introduced here may serve as a starting point for pre-compliance testing and for preparation for security certification.

    \section{Acknowledgments}
    We are grateful to Wenjamin Rosenfeld and Margarida Pereira for helpful discussions.
    This work was supported by the DFG under Germany's Excellence Strategy EXC-2111 390814868, the European project OpenQKD, the BMBF projects DE-QOR and QUBE-II, and the QuNET+ projects SKALE and MiQuE.
    AB acknowledges support by Elitenetzwerk Bayern in the PhD program ExQM.
    The funders played no role in study design, data collection, analysis and interpretation of data, or the writing of this manuscript.

    \section{Data availability}
    The source code used for data evaluation is available at~\cite{github_EmissionSecurityQKD} (MIT license).
    It includes the measurement pipeline (remote controlling the oscilloscope), the neural network, hyperparameter optimization routine, training pipeline and graphing of results.
    These materials can serve as a starting point for pre-compliance evaluation of other QKD devices.

    The measured data as recorded by the oscilloscope is available at~\cite{zenodo_measured_data}.
    The provided data and software allow to reproduce all reported results and may enable further work towards enhanced attacks via, e.g., improvements of the neural network.

%    \section{Author contributions}
%    AB and MS performed the experiments designed together with all other authors.
%    All authors contributed to preparing the manuscript.

    \clearpage

    \clearpage

    \newpage

	\appendix
    \onecolumngrid

	\begin{center}
	\Large{\textbf{Appendix}}
	\end{center}

    \section{Synchronization of Time Traces}
	\label{sec:appSynchronization}

    \begin{figure}[htbp]
        \centering
        \includegraphics[trim={10cm 20cm 10cm 20cm}, clip, width=0.35\textwidth]{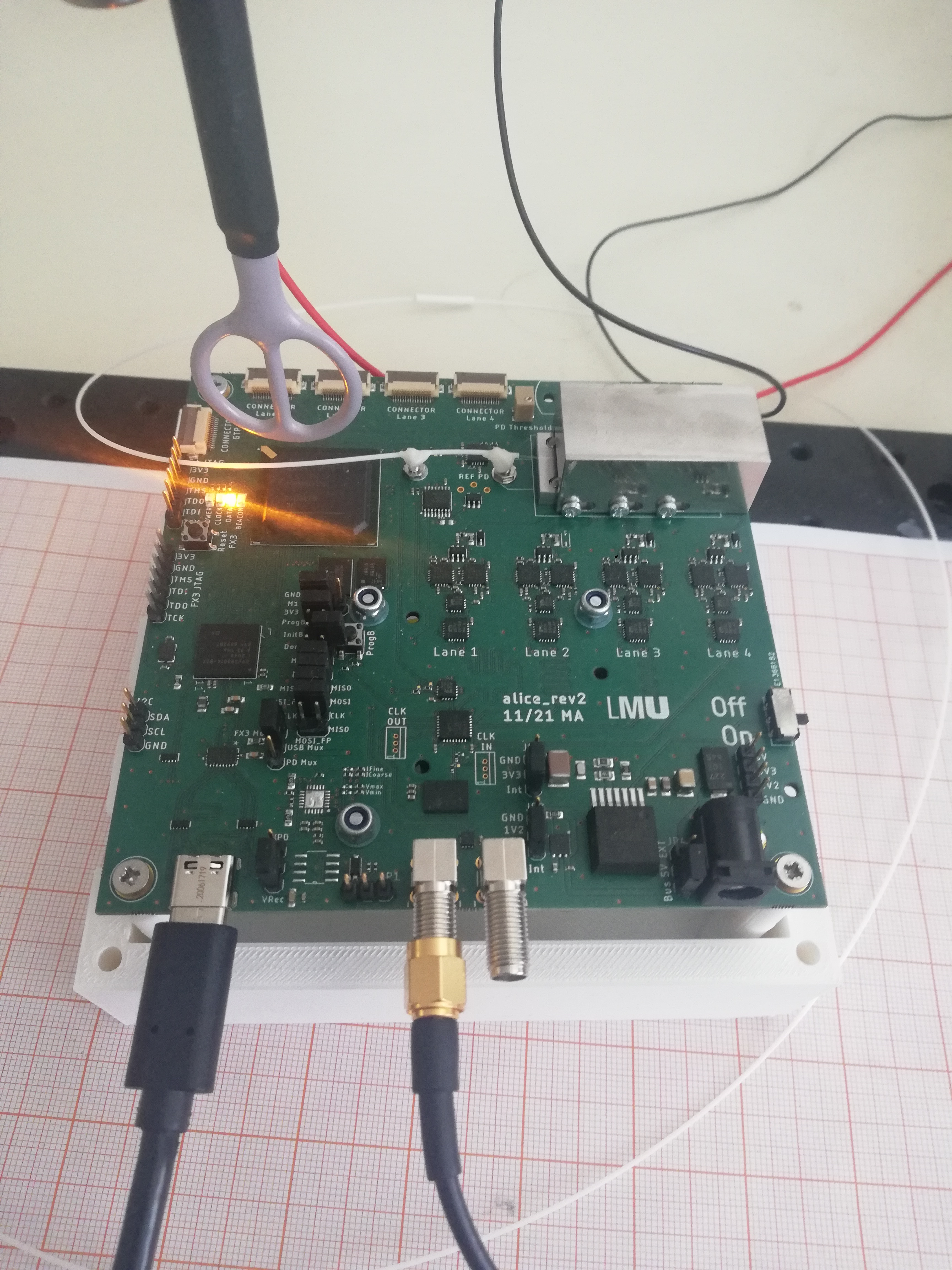}
        \caption{Photograph of the measurement setup for the revised electronics.
        The USB-C connection (bottom left) acts as a power supply and the SMA connection (bottom center) supplies the trigger signal for time synchronization.
        The fiber-connected optics module (top right) was switched off during our measurements in order to test the electronics.
        When active, its dedicated metal casing effectively shields rf emissions.
        }
        \label{fig:photo_nearfield_rev2}
    \end{figure}

    In order to predict the secret key from recorded emissions, it is necessary to synchronize the symbol sequence with the recorded emissions.
    We achieve this in two steps.
    First, we digitally obtain the clock signal from the recorded emissions via a narrow band-pass filter, allowing to synchronize the phase.
    I.e., we determine at which points (spaced 100 samples apart) in the recorded time trace a new symbol begins.
    Second, we synchronize the absolute time, i.e., find the point in the recorded time trace corresponding to the first symbol of the key.
    This is achieved using a separately recorded dedicated trigger signal from the device (See Fig.~\ref{fig:photo_nearfield_rev2}), which is always zero except for a short time indicating the start of the key.

    We ensure that recording the trigger signal does not affect our results by the following reference measurement.
    With otherwise identical setup, we record two sets of measurements, each set consisting of three datasets, namely training, validation (see App.~\ref{sec:appNetworkTraining}), and test.
    The first set of measurements is performed while recording the trigger signal for all datasets, which is used to synchronize and perform the attack as described in the main text.
    For the second set of measurements, we also record the trigger signal for training and validation datasets and use it for synchronization.
    However, we do not record the trigger signal for the test dataset.
    We observe that the training and validation accuracies between the two sets of measurements agree (within usual variation).

    For evaluating the test dataset of the second set of measurements, which does not contain a trigger signal, we only obtain the clock phase from the filtered probe signal, which leaves the absolute time of the first symbol to be determined.
    Even though lacking this information, we use the trained neural network to predict the key from the test dataset starting at an arbitrary symbol position.
    This results in a predicted sequence of symbols, which, in case of a perfect attack, reproduces the correct key symbols up to a circular shift.
    To evaluate the test accuracy, we correlate the true key with the predicted key, thus obtaining the optimal circular shift.
    In all cases, the optimal shift is unique and very easily distinguished from all other shifts.
    Using this optimal displacement, the test accuracy is consistent with that obtained for the first set of measurements, which include the trigger signal also in the test dataset.
    This shows that recording the trigger signal does not affect our results, justifying its use.

    For devices where no trigger signal is available at all, a different approach is needed to synchronize the training and validation datasets, since the above method only works for the test dataset.
    If the rf emissions are sufficiently strong, a header may be used within the key (e.g., a key section with a single repeated symbol, which leads to lower emissions during that time).
    The position of this header can be located within the measured data, thus providing synchronization of absolute time.
    For the training and validation datasets, this approach is consistent with the attacker model of Sec.~\ref{subsec:attacker-model}.
    For evaluating the test dataset, using a header in the key would violate the attacker model and the synchronization must be achieved by other means, e.g. by finding the optimal shift as described above.

    \section{Selective channel blocking attack}\label{sec:selective_channel_blocking}

    \begin{figure}[htb]
        \centering
        \includegraphics[width=0.95\textwidth]{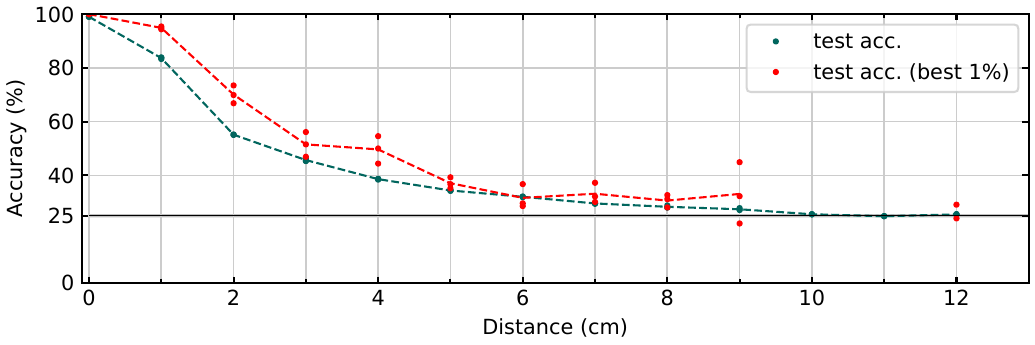}
        \caption{
		Same as Fig.~\ref{fig:distance} but also showing accuracy on a subset ($1\%$) of the test dataset.
        This subset is defined by selecting the 50 most confident (i.e., highest predicted probability) predictions for each symbol.
        The performance on the restricted test dataset is slightly better.
        For some locations ($10\,\mathrm{cm}$ and $11\,\mathrm{cm}$) the neural network never predicts a certain symbol.
        In that case, the selection procedure does not apply and test accuracy data on the restricted dataset is omitted.
        The high statistical variation across the test datasets is due to the much smaller size of the restricted dataset.
        }
        \label{fig:distance_best_percentile}
    \end{figure}

    A standard assumption in QKD is that the attacker has access to a lossless quantum channel.
    The levels of optical loss typical in QKD allow the attacker to very strongly preselect the optical pulses where the attack gives the most information and block all others.
    For example, if the attacker knows $3\%$ of the random key symbols with certainty and guesses the rest at random, this only achieves a raw key prediction accuracy of about $27.3\%$.
    However, at about $15\,\mathrm{dB}$ of optical loss, the attacker can block the $97\%$ unknown pulses, thus gaining complete knowledge of the sifted key.

    To select which pulses to block the attacker may use a model.
    In our case, the neural network outputs probabilities for each symbol (see App.~\ref{sec:appBitprediction}), which can be used directly.
    Assume that the attacker can afford to let through $1\%$ of pulses, which amounts to replacing $20\,\mathrm{dB}$ optical loss by a lossless channel.
    In order to not influence the proportions of symbols sent, the attacker should let through about the same number of pulses for each symbol.

    In simulating this scenario, we reevaluate the data of the distance measurement.
    The training procedure remains unchanged.
    For each test dataset consisting of $20\,000$ snippets, we artificially restrict to $1\%$ of snippets.
    We select $50$ snippets for each symbol, keeping only those where that symbol is predicted with the highest probability.
    While for our device and neural network, this only slightly improves the results (Fig.~\ref{fig:distance_best_percentile}), the possibility of selective channel blocking should be kept in mind when evaluating performance.
    This illustrates why it is not sufficient to only examine averaged performance metrics such as prediction accuracy.

    \section{Influence of basis choice and prediction of secret key bits}
	\label{sec:appBitprediction}

    \begin{table}[htb]
        \centering
        \caption{Confusion matrix of symbol predictions (test dataset) measured on the electronics without countermeasures.
        The data refers to the measurement shown in Fig.~\ref{fig:distance} at a distance of $1\,\mathrm{cm}$..
		The test accuracy of symbol prediction is $84.3\%$.
        }
        \label{tab:confusionmat}
        \renewcommand{\arraystretch}{2}
        \centering
        \begin{tabular}{ll|c|c|c|c|}
            \multicolumn{2}{c}{} & \multicolumn{4}{c}{Predicted Symbol} \\[-8pt]
            \multicolumn{2}{c}{} & \multicolumn{1}{c}{H} & \multicolumn{1}{c}{V} & \multicolumn{1}{c}{P} & \multicolumn{1}{c}{M}  \\
            \cline{3-6}
            \multirow{4}{*}{{\rotatebox[origin=c]{90}{True Symbol}
            }}
			& H & \textcolor[rgb]{0.004,0.4,0.37}{4917} & \textcolor[rgb]{0.55,0.32,0.04}{4} & \textcolor[rgb]{0.55,0.32,0.04}{5} & \textcolor[rgb]{0.55,0.32,0.04}{31} \\ \cline{3-6}
            & V & \textcolor[rgb]{0.55,0.32,0.04}{7} & \textcolor[rgb]{0.004,0.4,0.37}{3781} & \textcolor[rgb]{0.55,0.32,0.04}{813} & \textcolor[rgb]{0.55,0.32,0.04}{500}  \\ \cline{3-6}
            & P & \textcolor[rgb]{0.55,0.32,0.04}{6} & \textcolor[rgb]{0.55,0.32,0.04}{781}  & \textcolor[rgb]{0.004,0.4,0.37}{3860} & \textcolor[rgb]{0.55,0.32,0.04}{327}  \\ \cline{3-6}
            & M & \textcolor[rgb]{0.55,0.32,0.04}{9} &\textcolor[rgb]{0.55,0.32,0.04}{427}  & \textcolor[rgb]{0.55,0.32,0.04}{220}  & \textcolor[rgb]{0.004,0.4,0.37}{4299} \\ \cline{3-6}
        \end{tabular}
    \end{table}

    In the main text we evaluate the effectiveness of the attacks using prediction accuracy, i.e., the fraction of correctly recovered raw key symbols.
    There are several ways to evaluate information leakage and to put this quantity into perspective, e.g., for comparison with other side channel attacks.
    The neural network does not merely predict each symbol, but rather assigns to each snippet a probability distribution over all possible symbols (H, V, P, M), of which the most likely is taken as the prediction.
    These probabilities are used to calculate categorical cross-entropy, which is a measure of information leakage to the attacker and is minimized when training the neural network.

    Note that no single averaged metric is sufficient to rule out successful attacks.
    This is because an attacker can strongly pre-select the set of pulses to those leaking the most information and block the remaining optical pulses, see App.~\ref{sec:selective_channel_blocking}.

	Here, we want to relate the raw key prediction accuracy to the sifted key prediction accuracy.
    A standard assumption is that the attacker obtains access to the basis choices during the post-processing phase.
    In this case, there are two ways to evaluate the fraction of correctly recovered bits in the sifted key.

    First, the predicted symbols of the neural network can be represented as a \textit{confusion matrix}, which shows what symbols are more easily distinguishable than others, see Tab.~\ref{tab:confusionmat} for an example.
    By associating bit values to the symbols, the prediction accuracy for sifted key bits can be computed.
    A bit prediction is still correct when the network confuses two symbols that represent the same bit value in the sifted key, which leads to higher accuracies for bit prediction than symbol prediction.
    Assuming that the symbols H and P represent the bit value $0$ while V and M represent $1$, the bit prediction accuracy is $89.0\%$ for the example in Tab.~\ref{tab:confusionmat}.
	However, since the optics design is largely independent of the electronics design, one can also remap which laser driver line corresponds to which symbol.
    If the laser drivers originally used for the symbols H and V are rewired to represent bit value $0$, and P and M to represent $1$, the bit prediction accuracy is $87.1$\%.
    If H and M represent the $0$ and V and P represent $1$, the bit prediction accuracy is $92.5\%$.

    A second way to evaluate the sifted key bit prediction accuracy is to train a neural network for binary classification.
    In our case, the results agree very closely with those obtained from the confusion matrix approach.
	The confusion matrix approach not only gives the attacker information about Alice's basis choices, useful for additional attacks using optical measurements, it also does not require retraining for a new driver/symbol mapping as opposed to the binary classification network.
	
    \section{Neural Network and Training}
	\label{sec:appNetworkTraining}
	
	\begin{table}[hbt]
	\caption{Architecture of the neural network.}\label{tab:netarchitecture}
	\begin{tabular}{ >{\hangindent=1em}p{69mm}  c  r}
	\toprule
	Layer Type & Output Shape & Parameters \\ \midrule
	  Input & (500) & 0 \\
	  Gaussian Noise: $\sigma$ = 0.1 & (500) & 0 \\
	  Convolution (1D): GeLU \newline\hspace{1cm}(dilation rate = 1, kernel size = 3) & (500, 13) & 52 \\
	  Max Pooling (1D): size = 2 & (250, 13) & 0 \\
	  Spatial Dropout (1D): 25\%  & (250, 13) & 0 \\
	  Batch Normalization & (250, 13) & 52 \\
	  Convolution (1D): GeLU \newline\hspace{3mm}(dilation rate = 2, kernel size = 15) & (250, 118) & 23,128 \\
	  Max Pooling (1D): size = 1 & (250, 118) & 0 \\
	  Spatial Dropout (1D): 25\%  & (250, 118) & 0 \\
	  Batch Normalization & (250, 118) & 472 \\
	  Convolution (1D): GeLU \newline\hspace{3mm}(dilation rate = 4, kernel size = 5) & (250, 100) & 59,100 \\
	  Max Pooling (1D): size = 4 & (62, 100) & 0 \\
	  Batch Normalization & (62, 100) & 400 \\
	  Spatial Dropout (1D): 25\%  & (62, 100) & 0 \\
	  Flatten & (6200) & 0 \\
	  Dense: GeLU & (224) & 1,389,024 \\
	  Dense: Softmax & (4) & 900 \\ \midrule
					& & Total: 1,473,128 \\ \bottomrule
	\end{tabular}
	\end{table}

    The neural network accepts as input a time trace of $500$ samples (corresponding to $5$ symbols with $100$ samples each).
    We normalize the input data to have mean $0$ and standard deviation $1$ across the entire dataset.
    In order to track progress during training and examine generalization of the neural network, we use an additional \textit{validation dataset}, which is obtained by an independent measurement of a single trace ($20\,000$ symbols) using the same raw key as for the training datasets.
    The prediction accuracy evaluated on the validation dataset we call \textit{validation accuracy}.

    The neural network architecture is shown in Tab.~\ref{tab:netarchitecture}.
    One-dimensional convolutional filters identify specific patterns corresponding to switching between symbol combinations, while
	max pooling and batch normalization layers reduce complexity and increase speed and robustness of the net, respectively.
	Dropout layers help avoid overfitting and the flattening layer reshapes the tensor dimensions.
	Finally, two dense layers are used to classify the signal into one of four classes corresponding to the symbols H, V, P, M.
    For more details on the layers, see, e.g., Refs.~\cite{Schmidhuber2015,Alzubaidi2021}.
    We implement the neural network using the TensorFlow computing framework~\cite{tensorflow2015-whitepaper} and train it by minimizing categorical cross-entropy~\cite{Goodfellow-et-al-2016} using the Adam (Adaptive Moment Estimation) optimizer~\cite{Kingma2015AdamAM}.
    With this setup, training on seven measurements of 20\,000 symbols each takes around $5\,\mathrm{min}$ on an Nvidia A40 GPU\@.

	While the network architecture is chosen manually, the hyperparameters (e.g., convolution kernel sizes and noise levels) are found by hyperparameter optimization using Hyperband-based methods~\cite{JMLR:v18:16-558}.
	This optimization also effectively removes the second max pooling layer by setting its size to $1$.
    The hyperparameter optimization is performed on a dedicated, larger set of 30 measurements and repeated for both versions of the electronics (with and without countermeasures).
    There is no meaningful performance difference between the two hyperparameter sets on measurements from either device.
    Therefore, to simplify the evaluation, we use the same neural network (obtained by hyperparameter optimization on data measured from the electronics without countermeasures) for all reported findings

\begin{figure}[htb!]
        \centering
        \includegraphics[width=0.95\textwidth]{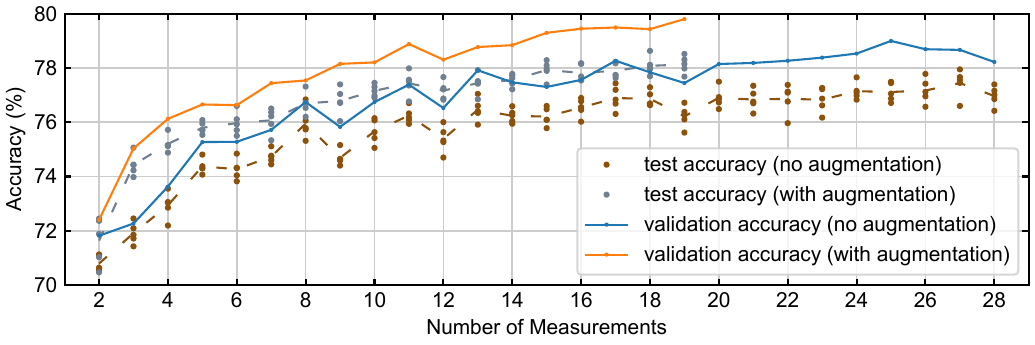}
        \caption{Dependence of validation and test accuracies on the amount of training and validation data.
		A single measurement (time trace) is always used for validation, the remaining are used for training.
        The network is trained independently for each number of measurements.
        Each trained neural network is evaluated on five independently measured test datasets.
        The test accuracies on each are shown separately (dots), as well as on average (dashed lines).
        This makes the analysis more robust against e.g. fluctuations in the noisy experimental environement.
        All data shown here is recorded above the FPGA at a distance of $2.5\,\mathrm{cm}$ from the circuit board.
		}
        \label{fig:trainingdata}
    \end{figure}

    We choose how much training data to collect at all locations using the larger dataset that is also used for hyperparameter optimization.
    The improvement of validation and test accuracy with increasing amount of measured training data and data augmentation is shown in Fig.~\ref{fig:trainingdata}.
    For data augmentation, we duplicate the traces and shift them by between one and three samples in either direction.
    The analysis suggests that seven measurements are representative of the attack's performance while keeping the data collection manageable.
	The test accuracies are evaluated on separate measurements of the same raw key, which implies a single-trace attack while also showing how reliable it is.

	Note that the discrepancy between validation and test accuracies is small.
    This indicates good generalization of the neural network and validates our approach of predicting the center symbol of a small snippet.
    In the revised electronics, we observed a slightly higher discrepancy between validation and test accuracies.

\end{document}